\documentclass[12pt]{article}
\usepackage{epsf}

\def\footnoteitem(#1)#2{
\begin{list}{#1}{\labelwidth4.0mm \leftmargin7.0mm
\labelsep2.5mm \rightmargin7.0mm \parsep0.5ex plus0.2ex minus0.1ex
\itemsep0ex plus0.2ex }
\item #2
\end{list}
}

\makeatletter
\newcommand{\contraction}[5][1ex]{%
  \mathchoice
    {\contraction@\displaystyle{#2}{#3}{#4}{#5}{#1}}%
    {\contraction@\textstyle{#2}{#3}{#4}{#5}{#1}}%
    {\contraction@\scriptstyle{#2}{#3}{#4}{#5}{#1}}%
    {\contraction@\scriptscriptstyle{#2}{#3}{#4}{#5}{#1}}}%
\newcommand{\contraction@}[6]{%
  \setbox0=\hbox{$#1#2$}%
  \setbox2=\hbox{$#1#3$}%
  \setbox4=\hbox{$#1#4$}%
  \setbox6=\hbox{$#1#5$}%
  \dimen0=\wd2%
  \advance\dimen0 by \wd6%
  \divide\dimen0 by 2%
  \advance\dimen0 by \wd4%
  \vbox{%
    \hbox to 0pt{%
      \kern \wd0%
      \kern 0.5\wd2%
      \contraction@@{\dimen0}{#6}%
      \hss}%
    \vskip 0.2ex%
    \vskip\ht2}}

\newcommand{\contraction@@}[3][0.06em]{%
  \hbox{%
    \vrule width #1 height 0pt depth #3%
    \vrule width #2 height 0pt depth #1%
    \vrule width #1 height 0pt depth #3%
    \relax}}
\makeatother

\def\secteq#1{ \setcounter{equation}{0}
\renewcommand{\theequation}{#1.\arabic{equation}} }

\begin{document}
%%%%%%%%%%%%%%%%%%%%%%%%%%%%%%%%%%%%%%%%%%%%%%%%%%%
\newcommand{\be}{\begin{equation}}
\newcommand{\ee}{\end{equation}}
\newcommand{\ba}{\begin{eqnarray}}
\newcommand{\ea}{\end{eqnarray}}

\newcommand{\cL}{{\cal L}}
\newcommand{\cM}{{\cal M}}
\newcommand{\Bt}{{\tilde B}}
\newcommand{\cO}{{\cal O}}
\newcommand{\cOt}{{\tilde\cO}}
\newcommand{\bt}{{\tilde\beta}}
\newcommand{\tr}{{\mbox{tr}\,}}
\newcommand{\str}{{\mbox{str}\,}}
\newcommand{\Exp}{{\mbox{exp}\,}}
\newcommand{\Mdot}{{\dot M}}
\newcommand{\Mbar}{{M_{VS}}}
\newcommand{\tb}{{\tilde\beta}}
\newcommand{\vp}{{\vec p}}
\newcommand{\hX}{{\hat X}}
\newcommand{\diag}{{\rm diag}}
\newcommand{\sbar}{{\overline{s}}}
\newcommand{\dbar}{{\overline{d}}}
\newcommand{\ubar}{{\overline{u}}}
\newcommand{\qbar}{{\overline{q}}}
\newcommand{\psibar}{{\overline{\psi}}}
\newcommand{\tu}{{\tilde u}}
\newcommand{\tub}{{\overline{\tu}}}
\newcommand{\td}{{\tilde d}}
\newcommand{\tdb}{{\overline{\td}}}
\newcommand{\ts}{{\tilde s}}
\newcommand{\tsb}{{\overline{\ts}}}
\newcommand{\ie}{{\it i.e.}}
\newcommand{\cf}{{\it cf.}}
\newcommand{\etc}{{\it etc.}}
\newcommand{\Nh}{{\hat N}}
\newcommand{\Real}{{\rm Re}}
\newcommand{\Imag}{{\rm Im}}
\newcommand{\eps}{{\varepsilon}}

\begin{titlepage}
\begin{flushright}
%preprint numbers\\
\end{flushright}
\vskip 1.8cm
\begin{center}
{\large \bf Quenched penguins and the $\Delta I=1/2$ rule}
\vskip1cm {Maarten Golterman$^a$ and  Elisabetta Pallante$^b$ }\\
\vspace{.5cm} {\small{\sl $^a$Department of Physics and Astronomy,
 San Francisco State University,\\ 
 1600 Holloway Ave, San Francisco, CA 94132, USA\\
\vspace{.2cm} $^b$Institute for Theoretical Physics, University of Groningen,
Nijenborgh 4,\\ 9747 AG  Groningen, The Netherlands  }}\\
 \vspace{2.0cm}

{\bf Abstract\\[10pt]} \parbox[t]{\textwidth}{{\small{
The transformation properties  of
strong penguin operators under the action of the flavor group change
when they are considered as operators in (partially) quenched QCD
instead of the unquenched theory.   As a result, additional
operators and new low-energy constants appear in the effective theory
describing non-leptonic kaon decay matrix elements in the partially quenched
setting.   These new low-energy constants do not have a counterpart in the
unquenched theory, and should thus be considered as an artifact
of the quenched approximation.
Here we consider strong penguin operators consisting of
products of two left-handed flavor currents, and give a complete one-loop
analysis in the effective theory for $K^0$ to $vacuum$ and $K^+$ to $\pi^+$
matrix elements.  We find that the new low-energy constants already
appear in these matrix elements at leading order.  This implies that
(partially) quenched lattice computations of for instance the
$\Delta I=1/2$ rule are affected by ambiguities intrinsic to the
use of the quenched approximation at leading order. The only exception
is the partially quenched case with three light sea quarks, consistent
with general expectations.  Our results are also relevant when
the charm quark is kept in the theory.}
}}

\end{center}
\vfill
{\small \noindent $^a$e-mail: { maarten@stars.sfsu.edu}  \\
\noindent $^b$e-mail: { e.pallante@rug.nl}  }\\
\end{titlepage}
\section{\large\bf Introduction}
\secteq{1}
Recently, there has been a renewed interest in lattice computations of
weak matrix elements relevant for the understanding of aspects of non-leptonic
kaon decays, such as the $\Delta I=1/2$ rule and $\varepsilon'/\varepsilon$, which
parametrizes direct CP violation in $K\to\pi\pi$ decays -- 
for a recent review, see
Ref.~\cite{review}.  However, while this renewed interest is due to the 
advent of
lattice fermions with very good chiral symmetry, to date all such computations
have been done in the quenched approximation, in which sea-quark effects
are ignored -- recent quenched results for both the $\Delta I=1/2$ rule and
$\varepsilon'/\varepsilon$ can be found in Refs.~\cite{cppacs,rbc}). 

In the theory where the charm quark has been integrated out penguin 
operators play
an important role.  In particular, referring to a commonly used basis, the
left-left (LL) penguin operator $\cO_2$ plays an important role in the
$\Delta I=1/2$ rule, while the left-right (LR) operator $\cO_6$  gives a 
major contribution to $\varepsilon'/\varepsilon$ \cite{dghb}.

When one makes the transition from unquenched QCD to partially-quenched (PQ)
QCD, the theory is changed from the physical theory with three light quarks
to a theory with $K$ light valence quarks and $N$ light sea quarks.
Fully quenched QCD is the special case with $N=0$.  This implies that the
flavor symmetry group changes from the usual $SU(3)_L\times SU(3)_R$ to
the graded group $SU(K+N|K)_L\times SU(K+N|K)_R$ \cite{bgpq}.
In general, this implies that the classification of weak operators with respect
to the flavor symmetry group also changes.  In particular, what happens for
strong penguins is that the penguin operator which transformed as a 
component of one irreducible representation (irrep) of $SU(3)_L\times SU(3)_R$
(the octet representation) now splits into several parts, each of which transforms
in a {\em different} representation of the PQ symmetry group.  One of those is the
``natural" generalization of the original penguin operator to the PQ theory,
whereas the other transforms in a more complicated way under
$SU(K+N|K)_L\times SU(K+N|K)_R$.  We will refer to these two parts as
the ``singlet" and ``adjoint" operators, respectively -- for reasons that will
become clear in Sec.~2.

At the level of the effective theory, this means that new low-energy constants (LECs)
occur for the adjoint operators, with no counterpart in the
unquenched theory, and which thus should be considered an artifact of the use
of an approximation where the number of light sea quarks is not equal 
to three.\footnote{We will see that for $N=3$, and when choosing valence-
and sea-quark masses equal, physical matrix elements of all new operators
vanish.}  These new LECs appear in physical matrix elements, unless one
decides to drop the corresponding adjoint operators from the PQ theory -- meaning
that the definition of the penguin operators themselves is changed in the
transition to the PQ theory.   We observe that also the LECs for the singlet
operators do not have to be equal to those of the physical three-flavor theory
if $N\ne 3$; even their scale dependence will in general be different.

This problem was considered in a previous paper for the strong LR operators
$\cO_{5,6}$ \cite{gp2}.  Here we consider the same problem for the LL operators
$\cO_{1,2}$.   While the group theory involved in the LL case is a little more
complicated than for the LR case, our results are rather similar.  We find that
also in the LL case the new adjoint operators do contribute to physical matrix 
elements already at the leading order (at tree level) in chiral perturbation 
theory (ChPT).\footnote{A speculation in Ref.~\cite{gp2} that 
they would only start contributing at order $p^4$ turns out to be incorrect.}
Here we demonstrate this explicitly with the examples of the $K^0$ to $vacuum$ and
$K^+$ to $\pi^+$ matrix elements.   We also find that the one-loop corrections
for the adjoint operators differ from those of the singlet operators,
calculated in Ref.~\cite{gp1}. In other words, the singlet and adjoint LECs do
not occur in some fixed, given linear combinations in physical matrix 
elements beyond tree level.

The outline of this paper is as follows.  In Sec.~2 we show how the standard
LL strong penguins break up into terms that transform differently under the
enlarged symmetry group of PQ QCD with $K$ valence quarks and $N$ sea quarks.
In Sec. 3~we construct the corresponding weak operators in ChPT, while Sec.~4
contains a discussion of some relevant peculiarities that arise for 
representations of the
graded symmetry group $SU(K+N|K)$ in the PQ theory.   Section~5 contains 
explicit expressions for the $K^0$ to $vacuum$ and $K^+$ to $\pi^+$ matrix elements
to one loop in ChPT.  Section~6 summarizes our results for the fully quenched
case, and the final section contains our conclusions.  Some group-theoretical
details and useful relations amongst weak operators are given in two 
appendices.
\section{\large\bf Left-Left penguins in partially quenched QCD}
\secteq{2}

We consider the LL penguin operators 
\ba
\label{penguins}
\cO_1&=&(\sbar d)_L (\ubar u)_L - (\sbar u)_L (\ubar d)_L\\
&=&(\sbar_\alpha d_\alpha)_L (\ubar_\beta u_\beta+\dbar_\beta d_\beta+\sbar_\beta s_\beta)_L -
(\sbar_\alpha d_\beta)_L (\ubar_\beta u_\alpha+\dbar_\alpha d_\beta+\sbar_\beta s_\alpha)_L\ ,
\nonumber\\
 \cO_2&=&(\sbar d)_L (\ubar u)_L + (\sbar u)_L (\ubar d)_L
 +2(\sbar d)_L (\dbar d+\sbar s)_L\nonumber\\
&=&(\sbar_\alpha d_\alpha)_L (\ubar_\beta u_\beta+\dbar_\beta d_\beta+\sbar_\beta s_\beta)_L +
(\sbar_\alpha d_\beta)_L (\ubar_\beta u_\alpha+\dbar_\alpha d_\beta+\sbar_\beta s_\alpha)_L\ ,
\nonumber
\ea
where
\be
\label{shorthand}
(\qbar_1 q_2)_L\equiv\qbar_1\gamma_\mu P_L q_2\ ,
\ee
with the projection operator on left-handed chirality $P_L=(1-\gamma_5)/2$.
In the second expression for each operator, we have made the color indices $\alpha,\beta$
explicit.  Both operators $\cO_{1,2}$ are penguin operators, and each is a linear combination of
color un-mixed and color mixed terms.  Both operators transform in the octet representation
of $SU(3)_L$, and, trivially, in the singlet representation of $SU(3)_R$.
%%%%%%%%%%%%%%%%%%
They are part of a basis of irreducible representations of the chiral 
group that are CPS invariant and with definite isospin $I=1/2$ and 
$I=3/2$~\cite{dghb}. In Appendix A we clarify how this basis is related to a 
set of weak operators more frequently used in phenomenological analyses. Our 
basis is especially convenient for working out group theoretical properties. 
%%%%%%%%%%%%%%%%%%%%%%%%%%%%%%%%%%%%%%%%%%%%%%%%%%%

\noindent As already mentioned in Sec.~1,
when we consider the LL penguin operators $\cO_{1,2}$ of Eq.~(\ref{penguins}) in the partially quenched theory, 
the representation content of these operators  changes.  
A general realization of PQ QCD contains $K$ valence
quarks, each accompanied by one of $K$ ghost quarks with the same mass
in order to suppress the valence-fermion determinant, and $N$ sea quarks -- the dynamical quarks -- which can all have masses
different from those of the valence quarks.   The relevant flavor symmetry 
group enlarges from the physical $SU(3)_L\times SU(3)_R$ to the graded group 
$SU(K+N|K)_L\times SU(K+N|K)_R$ \cite{bgpq}.\footnote{
For a detailed analysis of the actual symmetry group in the euclidean lattice theory, we refer
to Ref.~\cite{dgs}.  The upshot is that for our purposes, it is appropriate to
consider the PQ symmetry group to be $SU(K+N|K)_L\times SU(K+N|K)_R$
\cite{shsh}.}

It is clear that the $(\sbar d)_L$ factors of both operators in 
Eq.~(\ref{penguins})  are still a component of the
adjoint representation of $SU(K+N|K)$,%
\footnote{We will often drop the subscript $L$ 
on the group from now on, since all operators considered in this paper are
 trivial with respect to the right-handed group.} 
while the factors $(\ubar u+\dbar d+\sbar s)_L$ no longer
transform as singlets.  Instead, the operators can now be written as
\ba
\label{split}
\cO_1&=&\frac{K}{N}\;\cO_-^{PQS}+\cO_-^{PQA}\ ,\\
\cO_2&=&\frac{K}{N}\;\cO_+^{PQS}+\cO_+^{PQA}\ ,\nonumber\\
&&\nonumber\\
\cO_\pm^{PQS}&=&(\qbar_\alpha\Lambda q_\alpha)_L (\qbar _\beta q_\beta)_L
\pm (\qbar_\alpha\Lambda q_\beta)_L (\qbar _\beta q_\alpha)_L\ ,\nonumber\\
\cO_\pm^{PQA}&=&(\qbar_\alpha\Lambda q_\alpha)_L (\qbar _\beta A q_\beta)_L
\pm (\qbar_\alpha\Lambda q_\beta)_L (\qbar _\beta A q_\alpha)_L\ ,\nonumber
\ea
where we introduced the spurion fields $\Lambda$ and $A$ with values
\ba
\label{spurions}
\Lambda_i^{\ j}&=&\delta_{i3}\delta^{j2}\ ,\\
A&=&\diag\;\left(1-\frac{K}{N},\dots,-\frac{K}{N},\dots\right)\ .\nonumber
\ea
The first $K$ diagonal elements of $A$ are equal to $(1-K/{N})$ 
-- corresponding to the $K$ valence quarks -- and the last $N+K$ diagonal 
elements are equal to $(-K/N)$ -- corresponding to the $N$ sea quarks 
and the $K$ ghost quarks, both of which do not occur in $\cO_{1,2}$.  
The quark fields are graded vectors in flavor space, with fermionic components 
given by the valence and sea quarks, and bosonic components by ghost quarks.
The indices $i$ and $j$ are graded flavor indices,
and run over valence, sea and ghost flavors.\footnote{Wherever explicitly 
written, we denote flavor indices with $i,j,\ldots$, color indices with 
$\alpha , \beta ,\ldots$ and Dirac indices with $a,b,\ldots $.} 
For the down (strange) quark we have $i=2$
($i=3$).  The completely quenched theory, \ie\ the theory with $N=0$, for which the split of Eqs.~(\ref{split},\,\ref{spurions}) is singular, will be dealt with in Sec.~6.

We note at this point that for $N=K=3$ we regain the physical three-flavor 
theory. The adjoint operators $\cO_\pm^{PQA}$ now contain only terms involving 
either sea or ghost quarks, and it is rather straightforward to see that 
their contributions to physical matrix elements
(\ie\ those with only valence quarks on the external lines) vanishes because of
cancellation between sea-quark and ghost-quark loops. For this cancellation to
happen, valence masses and sea masses should be chosen equal.

The spurions $\Lambda$ and $A$ both transform in the adjoint representation 
of $SU(K+N|K)_L$, as can be seen from the fact that both have a vanishing 
supertrace (str) \cite{bb}.
The operators $\cO_\pm^{PQS}$ thus transform in the adjoint 
representation, while the operators $\cO_\pm^{PQA}$
transform as the product representation of two adjoint irreps, and they are 
thus reducible.
The corresponding decomposition of $\cO_\pm^{PQA}$ is accomplished by (anti-)symmetrization
in covariant and contravariant indices, and by ``removing" supertraces on 
pairs of covariant and contravariant indices, much as is done in the case of $SU(N)$ \cite{bb}.
Here we take the quark fields $q_i$ as covariant, and the anti-quark fields
$\qbar^i$ as contravariant.  It turns out that the operators $\cO^{PQA}_-$ and
$\cO^{PQS}_-$ ($\cO^{PQA}_+$ and $\cO^{PQS}_+$)
are already symmetric (anti-symmetric) in both their two covariant and their two contravariant 
flavor indices -- see Appendix B for details.  

In the next section, we are going to construct the low-energy bosonized 
effective lagrangian for the operators $\cO_{1,2}$ in the PQ theory. For this 
we do not need the details of the decomposition into irreps, and we  
postpone further discussion of this decomposition to Sec.~4.

\section{\large\bf The effective lagrangian}
\secteq{3}

The bosonized low-energy effective lagrangian is constructed in terms of the 
non-linear field
\be
\label{field}
\Sigma={\rm exp}{(2i\Phi/f)}\ ,
\ee
where $\Phi$ is the $(2K+N)\times (2K+N)$ hermitian field describing the Goldstone mesons,
and $f$ is the pion-decay constant in the chiral limit (normalized such that $f_\pi=132$~MeV).
Out of this field, we construct the necessary building blocks
\ba
\label{bblocks}
L_\mu&=&i\Sigma\partial_\mu\Sigma^\dagger\ ,\\
X_+&=&2B_0(\Sigma M^\dagger+M\Sigma^\dagger)\ , \nonumber
\ea
where $M$ is the $(2K+N)\times (2K+N)$ quark-mass matrix, and $2B_0=-\langle\ubar u\rangle/f^2$
in the chiral limit \cite{gl}.
These building blocks as well as the spurions $\Lambda$ and $A$ all transform in the
same way under the left-handed group $SU(K+N|K)$, and to lowest order in the chiral
expansion, we can construct the effective operators (note that $[\Lambda,A]=0$
and that\footnote{If we consider the PQ theory with the $\eta'$ integrated out.} 
$\str(L_\mu)=0$)
\ba
\label{effops}
\cL_1^A&=&\str(\Lambda L_\mu)\;\str(AL_\mu)\ ,\\
\cL_2^A&=&\str(\Lambda L_\mu AL_\mu)\ ,\nonumber\\
\cL_3^A&=&\str(\Lambda AL_\mu L_\mu)\ ,\nonumber\\
\cL_4^A&=&\str(\Lambda AX_+)\ .\nonumber
\ea
In deriving this list, we have also used $CPS$ symmetry \cite{cbetal}.  
For the singlet operators $\cO_\pm^{PQS}$ one replaces $A\to{\bf 1}$, and 
the above operators reduce to
\ba
\label{effopssinglet}
\cL_1^S&=&\str(\Lambda L_\mu L_\mu)\ ,\\
\cL_2^S&=&\str(\Lambda X_+)\ .\nonumber
\ea
The singlet operators $\cO_\pm^{PQS}$ are represented in chiral perturbation
theory by the leading order lagrangian \cite{cbetal,gp1}
\be
\label{Lsinglet}
\cL^S=-\alpha^{(8,1)}_1\;\cL_1^S+\alpha^{(8,1)}_2\;\cL_2^S\ ,
\ee
where $\alpha^{(8,1)}_{1,2}$ are two weak LECs.\footnote{
The minus sign is there to make the definition of $\alpha^{(8,1)}_1$ conform to
that of Ref.~\cite{cbetal}, in which the effective lagrangian was defined in
Minkowski space.  We work in euclidean space.}  Note that
the LECs for the two different operators $\cO_\pm^{PQS}$ are independent of 
each other,
even though we use the same symbol for both of them.

The operators $\cO_\pm^{PQA}$ correspond to different representations of
$SU(K+N|K)$, and are thus represented by different linear combinations of
$\cL_{1,2,3,4}^A$.  A mostly straightforward analysis leads to the bosonization 
rules\footnote{Some of the less 
straightforward aspects will be discussed in the next section.
See also Appendix B.}
\be
\label{bosonization}
\cO_\pm^{PQA}\to\cL^A_\pm= \alpha^{A\pm}_{1a}(\cL_1^A\pm\cL_2^A)
+\alpha^{A\pm}_{1b}\cL_3^A+\alpha^{A\pm}_2\cL_4^A\ .
\ee
Here we have explicitly indicated the dependence of the LECs on the
operator through the superscripts $\pm$, because they refer to different
representations of the PQ flavor group.  We conclude that the transition from
the unquenched theory to the PQ theory leads to the introduction of three
new LECs for each of the two operators $\cO_1$ and $\cO_2$.

\section{\large\bf Mixing of four-quark operators}
\secteq{4}

As stated already in the previous section, the two operators $\cO_\pm^{PQA}$
correspond to two different representations of $SU(K+N|K)$.  Both four-quark
operators may be written in the form $\qbar^i\qbar^j T_{ij}^{kl}q_k q_l$,
with $T$ symmetric (or anti-symmetric) in both $i\leftrightarrow j$ and
$k\leftrightarrow l$.\footnote{(Anti-)symmetrization here is understood as
appropriate for representations of graded groups.  ``Symmetrization" implies
symmetrization 
in bosonic indices and anti-symmetrization in fermionic indices, and
{\it vice versa} (see Appendix B and Ref.~\cite{bb}).}  Further decomposition of these representations
is generally possible by splitting the tensor $T$ into a part which is
supertraceless on any pair of covariant and contravariant indices, and a
supertrace part. Note that, because of the value of the spurion $\Lambda$, the
double supertrace of $T$ on both pairs of indices vanishes in our case.  

Taking the supertrace of $T$ on one pair, one obtains a tensor $S_i^k$, and one
may thus construct new four-fermion operators by replacing the tensor $T$
by the new tensor $S_i^k\delta_j^l$, of course after (anti-)symmetrizing this new tensor in
correspondence with the symmetry properties of the tensor $T$.
This results in the two new four-quark operators,
\ba
\label{newfq}
\cO^{PQT}_\pm&=&
(\qbar_\alpha\Lambda A q_\alpha)_L (\qbar _\beta  q_\beta)_L
\pm (\qbar_\alpha\Lambda  A q_\beta)_L (\qbar _\beta  q_\alpha)_L\\
&=&\left(1-\frac{K}{N}\right)\left(
(\sbar_\alpha d_\alpha)_L (\qbar _\beta  q_\beta)_L
\pm (\sbar_\alpha d_\beta)_L (\qbar _\beta  q_\alpha)_L
\right)\ .\nonumber
\ea
Our task is now to see whether these new operators are contained in the
original four-quark operators $\cO^{PQA}_\pm$, \ie\ whether the operators
$\cO^{PQA}_\pm$ and $\cO^{PQT}_\pm$ mix.   We observe that these operators
look just like the singlet operators $\cO^{PQS}_\pm$, but for the factor
$1-K/{N}$, which is the only remnant of the spurion $A$.  

There are three types of diagrams when one considers QCD
corrections to all these operators: self-energy corrections on the external 
lines,
vertex corrections, and penguin-like diagrams.  The mixing is only possible
through the penguin-like diagrams, because the first two types of diagrams
do not change the flavor structure of the operator.  This means that we have
to consider what happens when we contract one quark field with one
anti-quark field in each of these operators. Such a contraction,
in group-theoretical language, corresponds to taking a supertrace on 
a pair of quark and anti-quark indices of the tensor $T$.
Performing the contraction inside the operators $\cO^{PQA}_\pm$, we obtain\footnote{Wick contraction of graded quark bilinears gives 
$\contraction{}{\bar{q}}{^j_\beta}{q}
\bar{q}^j_\beta q_{i\alpha}= (-1)^{g(j)}\, S_{\alpha\beta}^j\,\delta^j_i$, with index $g(j)$ defined in Appendix B.}  
\ba
\label{Acontr}
\cO^{PQA}_\pm&\to& (\sbar_\alpha \gamma_\mu P_L d_\beta)
\Biggl(-\left(1-\frac{K}{N}\right)\left(\tr(\gamma_\mu P_LS^s_{\alpha\beta})+\tr(\gamma_\mu P_LS^d_{\alpha\beta})\right)
\\
&&\hspace{-1cm}
\mp\sum_{q\ {\rm valence}}\left(1-\frac{K}{N}\right)\tr(\gamma_\mu P_LS^q_{\alpha\beta})
\pm\sum_{q\ {\rm sea}}\frac{K}{N}\;\tr(\gamma_\mu P_LS^q_{\alpha\beta})
\mp\sum_{q\ {\rm ghost}}\frac{K}{N}\;\tr(\gamma_\mu P_LS^q_{\alpha\beta})
\Biggr)\nonumber\\
&&\pm(\sbar_\alpha \gamma_\mu P_L d_\alpha)
\Biggl(-\left(1-\frac{K}{N}\right)\left(\tr(\gamma_\mu P_LS^s_{\beta\beta})+\tr(\gamma_\mu P_LS^d_{\beta\beta})\right)
\nonumber\\
&&\hspace{-1cm}
\mp\sum_{q\ {\rm valence}}\left(1-\frac{K}{N}\right)\tr(\gamma_\mu P_LS^q_{\beta\beta})
\pm\sum_{q\ {\rm sea}}\frac{K}{N}\;\tr(\gamma_\mu P_LS^q_{\beta\beta})
\mp\sum_{q\ {\rm ghost}}\frac{K}{N}\;\tr(\gamma_\mu P_LS^q_{\beta\beta})
\Biggr)\ ,\nonumber
\ea
where $S^q_{\alpha\beta}$ is the quark propagator for flavor $q$ in an
arbitrary gluon background -- \ie\ the
quark propagator with an arbitrary number of gluons attached -- and we used 
that
\be
\label{lemma}
\gamma_\mu P_LS^q_{\alpha\beta}\gamma_\mu P_L=-\gamma_\mu P_L\;
\tr(\gamma_\mu P_LS^q_{\alpha\beta})\ .
\ee
The latter relation is easily proved by expanding the left-hand side on
a basis of $2^4$ euclidean hermitian Dirac matrices for fixed color 
indices $\alpha$ and $\beta$.
In these equations, tr stands for a trace over Dirac indices only.

Quark masses in our theory can be thought of as insertions leading to
higher-dimension weak operators, and we may thus consider the same result 
(\ref{Acontr}) in the chiral limit,
in which case the fermion propagator $S^q\to S$ becomes
flavor independent.  Equation~(\ref{Acontr}) thus simplifies to
\be
\label{Acontrsimple}
\cO^{PQA}_\pm\to-2\left(1-\frac{K}{N}\right)
\left((\sbar_\alpha\gamma_\mu P_L d_\beta)\;\tr(\gamma_\mu P_LS_{\alpha\beta})
\pm(\sbar_\alpha \gamma_\mu P_L d_\alpha)\;\tr(\gamma_\mu P_LS_{\beta\beta})\right)\ .
\ee
The above expression vanishes if we take $K=N$, consistent with the fact that the
operators $\cO^{PQT}_\pm$ vanish in this case (\cf\ Eq.~(\ref{newfq})).  When $K\ne N$,
this result shows that indeed the operators $\cO^{PQA}_\pm$ mix with $\cO^{PQT}_\pm$,
because the gluons attached to the fermion propagator couple to the flavor-singlet bilinear $\qbar\gamma_\mu P_L q$.

Performing the same contraction on the operators $\cO^{PQT}_\pm$, we find
\be
\label{Tcontrsimple}
\cO^{PQT}_\pm\to\left(1-\frac{K}{N}\right)(\mp N-2)
\left((\sbar_\alpha\gamma_\mu P_L d_\beta)\;\tr(\gamma_\mu P_LS_{\alpha\beta})
\pm(\sbar_\alpha\gamma_\mu P_L d_\alpha)\;\tr(\gamma_\mu P_LS_{\beta\beta})\right)\ .
\ee
Comparison of Eqs.~(\ref{Tcontrsimple}) and (\ref{Acontrsimple}) implies 
that there exist the supertraceless linear combinations
\be
\label{subtr}
\cO^{PQA}_\pm +\frac{2}{(\mp N-2)}\;\cO^{PQT}_\pm
\ee
that do not mix with the operators $\cO^{PQT}_\pm$.  In group-theoretical
language, these linear combinations are irreducible under $SU(K+N|K)$.  
We thus find that
in general, except for $K=N$, our four-quark operators do ``contain"
the operators $\cO^{PQT}_\pm$.  In the effective theory, the latter 
correspond to
the operator $\cL^A_3$ in Eq.~(\ref{effops}), justifying the bosonization
(\ref{bosonization}).   

A closer look at Eq.~(\ref{subtr}) reveals that when the number of sea quarks,
$N$, is equal to two, a singularity arises when one tries to decompose $\cO^{PQA}_-$
into irreps.  Clearly, for $N=2$ $\cO^{PQA}_-$ does mix with $\cO^{PQT}_-$, but
it is not possible to define an operator which does {\it not} mix with $\cO^{PQT}_-$.
In group theoretical language, the corresponding statement is that the 
representation in which $\cO^{PQA}_-$ transforms is not fully reducible.
This is indeed what one finds if one analyzes the tensor $T_{ij}^{kl}$ for this
operator, and in accordance with the fact that such reducible but not
decomposable reprentations do exist for graded groups \cite{bnrs}.\footnote{The operator
$\cO^{PQA}_-$ corresponds to a tensor which is symmetric in both
$i\leftrightarrow j$ and $k\leftrightarrow l$, with the appropriate grading
(see Appendix B).}  In any case, we still have to include the operator
$\cL^A_3$ in the effective theory, and the bosonization rule (\ref{bosonization})
thus stays the same.

\section{\large\bf $K^0\to$~$vacuum$ and $K^+\to\pi^+$ matrix elements}
\secteq{5}

In this section we give results for the simplest kaon matrix elements of the new
weak effective operators, $\cL^A_\pm$ (\cf\ Eq.~(\ref{bosonization})).  
We first give the tree-level results, and then
include also the chiral logarithms which occur at $O(p^4)$.  We will not 
present
a detailed analysis of $O(p^4)$ contact terms, because it is unlikely that the
matrix elements of $\cO^{PQA}_\pm$ will be numerically computed in the future.
The reason is that only the singlet LECs 
$\alpha^{(8,1)}_{1,2}$ are the interesting ones for physical predictions, 
and in a PQ setting they can be obtained
from the operators $\cO^{PQS}_\pm$, as long as the number of light sea quarks
is physical, \ie\ $N=3$ \cite{shshphys}.
For a complete $O(p^4)$ analysis of the singlet operators $\cO^{PQS}_\pm$,
represented at lowest order by $\cL^S$ in Eq.~(\ref{Lsinglet}), we refer to
Ref.~\cite{gp1} (see also Ref.~\cite{ls}). On the other hand, it is relevant 
to verify whether or not the tree level and one-loop contributions to 
physical matrix elements of $\cL^A$ have the same form as those arising from 
the singlet operator $\cL^S$.

At tree level, we find that
\ba
\label{Ktovactree}
\langle 0|\cL^A_\pm|K^0\rangle&=& \frac{4i(M_K^2-M_\pi^2)}{f}\;\left(1-\frac{K}{N}\right)
\;\alpha^{A\pm}_2\ ,\\
\langle \pi^+|\cL^A_\pm|K^+\rangle&=&
\frac{4M^2}{f^2}\;\left(1-\frac{K}{N}\right)\;(\mp\alpha^{A\pm}_{1a}-\alpha^{A\pm}_{1b}
-\alpha^{A\pm}_2)\ ,\nonumber
\ea
where in the case of $K^+\to\pi^+$ we took the kaon and pion masses to be equal,
$M_K=M_\pi=M$.  It is clear that the new LECs $\alpha^{A\pm}_{1a,1b,2}$ already
appear at leading order in ChPT, competing with the leading-order contributions
coming from the singlet LECs $\alpha^{(8,1)}_{1,2}$.

Next, we give the non-analytic terms arising at $O(p^4)$.  
We express our results in terms of bare meson masses,
where we take all sea quarks degenerate for simplicity, and also work in the
isospin limit, setting $m_u=m_d=m$, \ie\
\ba
\label{mesonmasses}
M_K^2&=&B_0(m+m_s)\ ,\\
M_\pi^2&=&2B_0 m\ ,\nonumber\\
M_{iS}^2&=&B_0(m_i+m_{\rm sea})\ ,\nonumber\\
M_{SS}^2&=&2B_0 m_{\rm sea}\ , \nonumber\\
M_{ij}^2&=&B_0 (m_i+m_j)\ , \nonumber
\ea
where $m_i$ is the mass of the $i$-th valence quark -- we use the labels
$i=u,d,s$ respectively $i=1,2,3$ interchangeably to label valence quark
masses.  For simplicity, we set $M_K=M_\pi=M$  in the $K^+\to\pi^+$ matrix 
elements, thus working in the degenerate limit.  It is further assumed 
that the $\eta'$ of the PQ 
theory is heavy, and therefore has been integrated out \cite{sh,gl1,shsh,gp1}.  We give
$\overline{MS}$ expressions for all one-loop results.

For the non-analytic terms in $\langle 0|\cO^{PQA}_\pm|K^0\rangle$ at one loop we find
\ba
\label{K0oneloop}
\langle 0|\cO^{PQA}_\pm|K^0\rangle_{\rm one-loop}&=&
\frac{4i\alpha^{A\pm}_{1a}}{f(4\pi f)^2}\Biggl\{
\left(1-\frac{K}{N}\right)M_{33}^4\left (1-\log{\frac{M_{33}^2}{\mu^2}}\right )
\nonumber\\&&\hspace{-4cm}
+ \frac{1}{N} \sum_{i\ {\rm valence}} \frac{M_{ii}^2-M_{SS}^2}
{M_{33}^2-M_{ii}^2}\left [ -M_{33}^4\left (1-\log{\frac{M_{33}^2}{\mu^2}}\right )+M_{ii}^4 \left (1-\log{\frac{M_{ii}^2}{\mu^2}}\right ) \right] 
\nonumber\\
&&\hspace{-4cm}
\pm\sum_{i\ {\rm valence}} M_{3i}^4 \left (1-\log{\frac{M_{3i}^2}{\mu^2}}\right ) \pm \frac{1}{N}\left(1-\frac{K}{N}\right) M_{33}^2 (3M_{33}^2-2M_{SS}^2)
\log{\frac{M_{33}^2}{\mu^2}}
\nonumber\\
&&\hspace{-4cm}
\mp K M_{3S}^4\left(1- \log{\frac{M_{3S}^2}{\mu^2}}\right)
\mp \frac{1}{N}\left(1-\frac{K}{N}\right) 
M_{33}^2 (2M_{33}^2-M_{SS}^2) 
- (m_3\leftrightarrow m_2)\Biggr\}
\nonumber\\
&&\hspace{-4cm}
+\frac{4i\alpha^{A\pm}_{1b}}{f(4\pi f)^2}\left(1-\frac{K}{N}\right)
\Biggl\{  \frac{1}{N} M_{33}^2\, (3M_{33}^2-2M_{SS}^2) \log{\frac{M_{33}^2}{\mu^2}}+N M_{3S}^4 
\left(1-\log{\frac{M_{3S}^2}{\mu^2}}\right) 
\nonumber\\
&&\hspace{-4cm}
- \frac{1}{N} M_{33}^2(2M_{33}^2-M_{SS}^2) -
(m_3\leftrightarrow m_2)\Biggr\}
\nonumber\\
&&\hspace{-4cm}
+\frac{2i\alpha^{A\pm}_2}{f(4\pi f)^2}(M_K^2-M_\pi^2)
\left(1-\frac{K}{N}\right)\Biggl\{ 
\frac{1}{N}\left [2M_{33}^2-M_{SS}^2+2M_{33}^2\frac{M_{SS}^2-M_{33}^2}
{M_{22}^2-M_{33}^2}\right ]\, \log{\frac{M_{33}^2}{\mu^2}} 
\nonumber\\
&&\hspace{-4cm}
+\frac{1}{N}(-3M_{33}^2+M_{SS}^2)+N M_{3S}^2\left(1
-\log{\frac{M_{3S}^2}{\mu^2}} \right)
+ (m_3\leftrightarrow m_2)\Biggr\}\, ,
\ea
where $m_3\leftrightarrow m_2$ stands for the exchange of strange and down 
valence-quark masses, and we took the exact isospin limit.

For the non-analytic terms in $\langle\pi^+|\cO^{PQA}_\pm|K^+\rangle$ at one 
loop, with $M_K=M_\pi=M$ and $M_{VS}^2=(M^2+M_{SS}^2)/2$, we find
\ba
\label{Kpioneloop}
\langle\pi^+|\cO^{PQA}_\pm|K^+\rangle_{\rm one-loop}&=&
\frac{4M^2\alpha^{A\pm}_{1a}}{f^2(4\pi f)^2}\Biggl\{ 
\left [\frac{2K}{N} M_{SS}^2 + \left (4-\frac{6K}{N}\right ) M^2\right ]
\log{\frac{M^2}{\mu^2}}+\frac{K}{N} M_{SS}^2 
\nonumber\\
&&\hspace{-4cm}
-\left (2-\frac{K}{N}\right ) M^2 
\pm (K-2N)M_{VS}^2\left(1-\log{\frac{M_{VS}^2}{\mu^2}}\right)
\pm M^2\left ( K+\frac{4}{N}  \left (1-\frac{K}{N}\right )\right ) 
 \nonumber\\
&&\hspace{-4cm}
\pm \frac{2}{N}\left (1-\frac{K}{N}\right ) M_{SS}^2
\pm\left [ 
\frac{6}{N}\left (1-\frac{K}{N}\right )M_{SS}^2
-M^2\left (K + \frac{16}{N }\left (1-\frac{K}{N}\right )  \right )
\right ]  \log{\frac{M^2}{\mu^2}}
  \Biggr\}\nonumber\\
&&\hspace{-4cm}
+\frac{4M^2\alpha^{A\pm}_{1b}}{f^2(4\pi f)^2}
\left(1-\frac{K}{N}\right)\Biggl\{ 
-N M_{VS}^2\left(1-\log{\frac{M_{VS}^2}{\mu^2}}\right)
+\frac{2}{N}\left (3M_{SS}^2-8M^2\right ) \log{\frac{M^2}{\mu^2}}
\nonumber\\
&&\hspace{-4cm}
+\frac{2}{N}M_{SS}^2+\frac{4}{N}M^2
\Biggr\}
\nonumber\\
&&\hspace{-4cm}
+\frac{4M^2\alpha^{A\pm}_{2}}{f^2(4\pi f)^2}
\left(1-\frac{K}{N}\right)\frac{2}{N}\Biggl\{ M_{SS}^2+(M_{SS}^2-4M^2) \log{\frac{M^2}{\mu^2}}
\Biggr\}\, .
\ea
%%%%%%%%%%%%%%%%
Eqs.~(\ref{K0oneloop}) and (\ref{Kpioneloop}) represent the partially quenched 
one-loop contamination which affects $K$ to $vacuum$ and $K^+$ to $\pi^+$ 
amplitudes through the operators $\cO_\pm^{PQA}$. By construction, these 
contributions 
vanish in the full QCD limit, when $N=K$ and the valence- and sea-quark masses 
are chosen to be pairwise equal.  We note that in Eq.~(\ref{Kpioneloop}) $M$ denotes
the mass of a meson made out of two valence quarks, \ie\ $M=M_{ij}$ for all $i$, $j$
valence.   The contribution of sea-quarks can be traced through the index $S$
on $M_{VS}$, which is the mass of a meson made out of one valence and one sea
quark, and $M_{SS}$, which is the mass of a meson made out of two sea quarks.
\section{\large\bf The quenched case}
\secteq{6}

In this section we summarize the results for the quenched case, \ie\ the case without
any sea quarks, for which the symmetry group is $SU(K|K)$.  The quenched
case needs to be treated separately, because the $\eta'$ cannot be integrated out
\cite{bgq,sh92}.  Moreover, the phenomenon that not all reducible representations
of graded groups are fully reducible already shows up when one tries to construct the
adjoint representation: the product of the fundamental irrep and its complex conjugate
is not fully reducible \cite{gp2}.  The decomposition (\ref{split}) now takes the form
\ba
\label{qsplit}
\cO_1&=&\frac{1}{1+\gamma}\left(\gamma\cO_-^{QS}+\cO_-^{QNS}\right)\ ,\\
\cO_2&=&\frac{1}{1+\gamma}\left(\gamma\cO_+^{QS}+\cO_+^{QNS}\right)\ ,\nonumber\\
\cO_\pm^{QS}&=&(\qbar_\alpha\Lambda q_\alpha)_L (\qbar _\beta q_\beta)_L
\pm (\qbar_\alpha\Lambda q_\beta)_L (\qbar _\beta q_\alpha)_L\ ,\nonumber\\
\cO_\pm^{QNS}&=&(\qbar_\alpha\Lambda q_\alpha)_L (\qbar _\beta \Nh q_\beta)_L
\pm (\qbar_\alpha\Lambda q_\beta)_L (\qbar _\beta \Nh q_\alpha)_L\ ,\nonumber
\ea
where the spurion field $\Nh$ is given by
\be
\label{qspurion}
\Nh=\diag\;\left(1,\dots,-\gamma,\dots\right)\ .\nonumber
\ee
The first $K$ diagonal elements of $\Nh$ are equal to $1$, and the last $K$ diagonal elements
are equal to $-\gamma$, with $\gamma$ arbitrary except 
$\gamma\ne -1$.\footnote{In Ref.~\cite{gp2} we chose $\gamma=1$.}
Note that $\Nh$ has a non-vanishing supertrace, unlike $A$ in the PQ case, while the
supertrace of the unit matrix $\diag(1,\dots,1)$ does vanish, again unlike 
the PQ case.
The operators $\cO^{QS}_\pm$ and $\cO^{QNS}_\pm$ each are represented by their own
LECs in the effective theory, just as in the case of LR penguins \cite{gp2}.
  Naively, it looks
like one could avoid those LECs which correspond to $\cO^{QS}_\pm$ by choosing
$\gamma=0$.  However, the operators $\cO^{QNS}_\pm$ mix with $\cO^{QS}_\pm$
{\textit{for any choice of $\gamma$}},
as can be shown in the quark language using arguments similar to those
of Sec.~4.  
 
Once again, in order to analyze results from quenched QCD, one has to develop 
the effective theory for the operators $\cO^{QNS}_\pm$.  This is 
straightforward: one exactly follows
the construction of Sec.~3 for the PQ case, by replacing the PQ spurion $A$ everywhere
with $\Nh$.  However, there is one extra operator, because in the quenched 
case $\str(L_\mu)\ne 0$ -- due to the $\eta^\prime$, which cannot be integrated out in the
quenched case --
and the quenched non-singlet effective lagrangian replacing 
Eq.~(\ref{bosonization})
reads now (see also Appendix B)
\ba
\label{LQNS}
\cL^{QNS}_\pm&=&\alpha^{N\pm}_{1a}\left(\str(\Lambda L_\mu)\;\str(\Nh L_\mu)\pm
\str(\Lambda L_\mu \Nh L_\mu)\right)\\
&&+\alpha^{N\pm}_{1b}\left(\str(\Lambda\Nh L_\mu L_\mu)\pm
\str(\Lambda\Nh L_\mu)\;\str(L_\mu)\right)+\alpha^{N\pm}_2\str(\Lambda\Nh X_+)\ .
\nonumber\ea

The tree-level $K^0$ to $vacuum$ and $K^+\to\pi^+$ matrix elements are given by
\ba
\label{qKtovactree}
\langle 0|\cL^N_\pm|K^0\rangle&=& \frac{4i(M_K^2-M_\pi^2)}{f}
\;\alpha^{N\pm}_2\ ,\\
\langle \pi^+|\cL^N_\pm|K^+\rangle&=&
\frac{4M^2}{f^2}\;(\mp\alpha^{N\pm}_{1a}-\alpha^{N\pm}_{1b}
-\alpha^{N\pm}_2)\ ,\nonumber
\ea
where we labeled the non-singlet effective operators and LECs with a superscript $N$
instead of a superscript $A$ for the quenched case.

Before presenting quenched one-loop results, we remind the reader 
that the tree-level propagator of a neutral valence meson
made out of quark and anti-quark flavor $i$ is given by
\be
\label{qX}
D_{ij}(p)=\frac{\delta_{ij}}{p^2+M_{ii}^2}-\frac{1}{3}\;\frac{m_0^2+\alpha p^2}{(p^2+M_{ii}^2)(p^2+M_{jj}^2)}\ ,
\ee
where $m_0^2$ is the ``double-hairpin" vertex at zero momentum, and $\alpha$
parametrizes its momentum dependence \cite{bgq}.
For the non-analytic terms at one loop, we find
\ba
\label{Ktovacq}
\langle0|\cO^{QNS}_\pm|K^0\rangle_{\rm one-loop}&=&
\frac{4i\alpha^{N\pm}_{1a}}{f(4\pi f)^2}\Biggl\{ M_{33}^4\left(1 - 
\log{\frac{M_{33}^2}{\mu^2}} \right)
-\frac{1}{3}(1+\gamma)\sum_{i\ {\rm valence}}\frac{1}{M_{33}^2-M_{ii}^2}\times
\nonumber\\&&\hspace{-4cm}
\left [ M_{ii}^4\left( m_0^2 -\alpha M_{ii}^2\right)
 \left (1-\log{\frac{M_{ii}^2}{\mu^2}}\right ) 
-M_{33}^4\left(m_0^2-\alpha M_{33}^2\right)\left (1-\log{\frac{M_{33}^2}{\mu^2}}\right )    
\right ]
\nonumber\\
&&\hspace{-4cm}
\pm  \frac{1}{3}M_{33}^2 ( m_0^2 - 2\alpha M_{33}^2)
\mp M_{33}^2\left (\frac{2}{3}m_0^2 -\alpha M_{33}^2\right )\,
\log{\frac{M_{33}^2}{\mu^2}} 
\nonumber\\
&&\hspace{-4cm}
\pm (1+\gamma ) \sum_{i\ {\rm valence}} M_{3i}^4 
\left (1-\log{\frac{M_{3i}^2}{\mu^2}}\right ) 
-\,(m_3\leftrightarrow m_2)\Biggr\}
\nonumber\\
&&\hspace{-4cm}
+\frac{4i\alpha^{N\pm}_{1b}}{f(4\pi f)^2} \Biggl\{
\left( \frac{1}{3}m_0^2 - \frac{2}{3}\alpha M_{33}^2\right)M_{33}^2 
-\left (\frac{2}{3}m_0^2 -\alpha M_{33}^2\right )\, M_{33}^2
\log{\frac{M_{33}^2}{\mu^2}} 
\nonumber\\
&&\hspace{-4cm}
\pm  M_{33}^4 
\left (1-\log{\frac{M_{33}^2}{\mu^2}}\right ) 
-\, (m_3\leftrightarrow m_2)\Biggr\}
\nonumber\\
&&\hspace{-4cm}
+\frac{2i\alpha^{N\pm}_2}{3f(4\pi f)^2}
(M_K^2-M_\pi^2)\Biggl\{ m_0^2 -3\alpha M_{33}^2 +
\left [-m_0^2+2\alpha M_{33}^2-2M_{33}^2
\frac{ m_0^2 -\alpha M_{33}^2}
{M_{33}^2-M_{22}^2} \right ]\,
\log{\frac{M_{33}^2}{\mu^2}} 
\nonumber\\
&&\hspace{-4cm}
+\, (m_3\leftrightarrow m_2)\Biggr\}\, ,
\ea
for the $\langle0|\cO^{QNS}_\pm|K^0\rangle$ matrix element, and
\ba
\label{Ktopiq}
\langle\pi^+|\cO^{QNS}_\pm|K^+\rangle_{\rm one-loop}&=&
\frac{4M^2\alpha^{N\pm}_{1a}}{3f^2(4\pi f)^2}  \Biggl\{
2(6M^2+K(1+\gamma ) (m_0^2-3\alpha M^2)) \log{\frac{M^2}{\mu^2}}
\nonumber\\
&&\hspace{-4cm}
-6M^2+ K(1+\gamma ) (m_0^2+\alpha M^2)
\pm (2m_0^2+(4\alpha +3K(1+\gamma ))M^2)
\nonumber\\
&&\hspace{-4cm}
\pm \left ( 6m_0^2-(16\alpha +3K(1+\gamma ))M^2\right )\,
 \log{\frac{M^2}{\mu^2}}
\Biggr\}
\nonumber\\
&&\hspace{-4cm}
-\frac{8M^2\alpha^{N\pm}_{1b}}{3f^2(4\pi f)^2}\Biggl\{
-(3m_0^2-8\alpha M^2) \log{\frac{M^2}{\mu^2}}-m_0^2-2\alpha M^2 
\pm 3M^2\left ( 1-2  \log{\frac{M^2}{\mu^2}}  \right )
\Biggr\}
\nonumber\\
&&\hspace{-4cm}
+\frac{8M^2\alpha^{N\pm}_{2}}{3f^2(4\pi f)^2}\Biggl\{ m_0^2+(m_0^2-4\alpha M^2)  \log{\frac{M^2}{\mu^2}}
\Biggr\}
\ea
for the $\langle\pi^+|\cO^{QNS}_\pm|K^+\rangle$ matrix element.

\section{\large\bf Conclusions}
\secteq{7}

We add a few comments to the main conclusions already stated in the Introduction.

First, a consistency check on all our explicit tree-level and one-loop results is that
they should vanish when the number of valence quarks $K$ is chosen equal to
the number of sea quarks $N$ and when the valence- and sea-quark masses
are also chosen to be pairwise equal.  We noted this already in Sec.~2, and it is
straightforward to see that indeed all results contained in Sec.~5 do satisfy
this requirement.

We emphasize that the new adjoint operators $\cO^{PQA}_\pm$ occurring in the 
PQ theory
are genuinely new operators, and one thus expects that one-loop corrections in
ChPT for matrix elements of these operators differ from those of the singlet
operators $\cO^{PQS}_\pm$.  We find that this is indeed the case.  If one however only considers the
tree-level results, only one particular linear combination of singlet and adjoint
LECs appears in all matrix elements (including $K\to\pi\pi$ matrix elements).
This is unlike the case of LR penguins, where enhancement of the adjoint
operators leads to the appearance of chiral logarithms already at leading order
in ChPT \cite{gp2}.  Thus, Refs.~\cite{cppacs,rbc}, while aiming for the 
leading-order
LECs  for $\cO_1$ and $\cO_2$
in the quenched approximation, have actually computed the linear combinations
of LECs that appear at tree level, namely
\ba
\label{comb}
&&\frac{1}{1+\gamma}\left(\gamma\alpha^{(8,1)}_2+\alpha^{N\pm}_2\right)
\ ,\\
&&\frac{1}{1+\gamma}\left(\gamma\alpha^{(8,1)}_1
\mp\alpha^{N\pm}_{1a}-\alpha^{N\pm}_{1b}-
(\gamma\alpha^{(8,1)}_2+\alpha^{N\pm}_2)\right)
\ .\nonumber
\ea
At first, the appearance of the arbitrary parameter $\gamma$ looks a bit 
strange.
However, this just reflects the fact that the non-singlet operators $\cO^{QNS}_\pm$
are only defined modulo mixing with the singlet operators  $\cO^{QS}_\pm$, since
they do not by themselves constitute an irrep, as discussed in Sec.~6.  In the PQ case
no such arbitrariness arises.

%%%%%%%%%%%%%%%%%%%%%%%%%%%%%%%%%
%We have also demonstrated, with the calculation of $K$ to $vacuum$ and $K^+$ to $\pi^+$ matrix %elements, that the one-loop corrections due to the new adjoint 
%operators differ from those of the singlet physical operators, calculated in 
%Ref.~\cite{gp1}. One important consequence is that the singlet and adjoint LECs do not occur in some %fixed, given linear combinations in physical matrix elements, beyond tree level.
%In other words, the new (partially) quenched LECs {\em cannot} be absorbed 
%into a redefinition of the physical LECs; this is true also at tree level (leading chiral order) where %quenched LECs already appear.  

Our results show that quenching artifacts do modify, already at the leading chiral order, those $\Delta S=1$ weak 
matrix elements that receive contributions from LL penguin operators.\footnote{We disagree 
with the conclusion of Ref.~\cite{Laiho} that the 
$\Delta I=1/2$ rule is not affected by (partial) quenching artifacts at tree 
level.  In particular, we point out that it is not possible to decide unambiguously within the quenched
approximation what is the ``best" linear combination of singlet and non-singlet
operators to choose.  Only an unquenched computation can, in hindsight, decide this issue. }
This is especially the case for the $\Delta I=1/2$ rule where the dominant 
contributions come from the current-current operators $Q_1$ and $Q_2$ of 
Eq.~(\ref{Qops}).
In addition, quenching contaminations to LL penguin operators can in principle 
affect a lattice determination of $\varepsilon '/\varepsilon$.\footnote{We thank the referee for pointing this out.}
While quenching artifacts analyzed in Ref.~\cite{gp2} affect a dominant contribution to $\varepsilon '/\varepsilon$ coming from the LR penguin 
operator $Q_6$, the quenching ambiguity affecting the operator $Q_4$ 
through $\cO_1$ and $\cO_2$ 
can be relevant in the presence of a large cancellation of the dominant contributions from $Q_6$ and the electroweak penguin operator 
$Q_8$~\cite{Japan}.

%%%%%%%%%%%%%%%%%%%%%%%
Finally, we note that our results are also relevant for the theory in which the charm
quark is kept.  In that case, the relevant weak operators are $\cO_\pm$, which can
be written as
\ba
\label{charmin}
\cO_-&=&(\sbar d)_L (\ubar u)_L - (\sbar u)_L (\ubar d)_L-(u\to c)\\
&=&\cO_1-(u\to c)\ ,\nonumber\\
\cO_+&=&(\sbar d)_L (\ubar u)_L + (\sbar u)_L (\ubar d)_L-(u\to c)\nonumber\\
&=&\frac{1}{5}\cO_2+\frac{2}{15}\cO_3+\frac{2}{3}\cO_4-(u\to c)\ ,\nonumber
\ea
where $\cO_{3,4}$ transform in the 27-dimensional irrep of $SU(3)_L$ -- 
see also Appendix A.  
The operators $\cO_{1,2}$ appear, and the discussion
of this paper thus also applies to the operators $\cO_\pm$.  The charm quark
transforms as a singlet under both $SU(3)$ and $SU(K+N|K)$, and the
``$(u\to c)$" terms in Eq.~(\ref{charmin}) thus transform in the adjoint 
representations
of both $SU(3)$ and $SU(K+N|K)$.  Our observations here do not apply to the
$SU(4)$ case of an unphysically light
charm quark, because the $SU(4)$ transformation properties of $\cO_\pm$ 
are different.

\section*{\large\bf Acknowledgements}

We would like to thank Norman Christ, Jack Laiho, 
Ruth van de Water and Giovanni Villadoro  for useful discussions.  
MG would like to thank the Institute for Theoretical Physics at the
University of Groningen, the Physics Department of the University of Rome ``La Sapienza,"
the Saha Institute for Nuclear Research in Kolkata, 
and the Nuclear Theory group
at Lawrence Berkeley National Laboratory for hospitality.
MG is supported in part by the US Department of Energy. 
\vspace{1.5truecm}
\section*{\large\bf Appendix A: $\Delta S=1$ weak operators}
\secteq{A}
In this appendix, we clarify the relation between the  operator basis we used and 
a set of weak operators often used in phenomenological 
analyses of kaon decays  -- a comprehensive review can be 
found in Ref.~\cite{buras}. 
Our basis of $\Delta S=1$ four-quark effective weak operators corresponds to
irreducible representations of the chiral group $SU(3)_L\times SU(3)_R$, that 
are CPS invariant \cite{cbetal} (\ie\, invariant under the product of CP 
and the exchange of strange and d quark), and with definite isospin, $I=1/2$ and $I=3/2$:
\ba
\label{basis}
(8_L,1_R)~~ I=1/2:~ 
\cO_1&=&(\sbar d)_L (\ubar u)_L - (\sbar u)_L (\ubar d)_L\, ,\\
I=1/2:~
\cO_2&=&(\sbar d)_L (\ubar u)_L + (\sbar u)_L (\ubar d)_L
+2(\sbar d)_L (\dbar d+\sbar s)_L\, ,\nonumber\\
(27_L,1_R)~~I=1/2:~
\cO_3&=&(\sbar d)_L [(\ubar u)_L+2(\dbar d)_L-3(\sbar s)_L]
+  (\sbar u)_L (\ubar d)_L\, ,\nonumber\\
I={3}/{2}:~
\cO_4&=& (\sbar d)_L [(\ubar u)_L-(\dbar d)_L]
+  (\sbar u)_L (\ubar d)_L\,  ,\nonumber\\
(8_L,1_R)~~I={1}/{2}:~
\cO_5&=&(\sbar d)_L [(\ubar u)_R+(\dbar d)_R+(\sbar s)_R]\, ,
\nonumber\\
I={1}/{2}:~
\cO_6&=&(\sbar_\alpha d_\beta )_L [(\ubar_\beta u_\alpha )_R+
(\dbar_\beta d_\alpha )_R+(\sbar_\beta s_\alpha )_R]\, ,\nonumber\\
(8_L,8_R)~~I={1}/{2}:~
\cO_7&=&(\sbar d)_L [(\ubar u)_R-(\sbar s)_R] - (\sbar u)_L (\ubar d)_R\ \, ,
\nonumber\\
I={1}/{2}:~
\cO_8&=&(\sbar_\alpha d_\beta)_L [(\ubar_\beta u_\alpha )_R-
(\sbar_\beta s_\alpha )_R]  - (\sbar_\alpha u_\beta)_L 
(\ubar_\beta d_\alpha )_R \, ,\nonumber\\
(8_L,8_R)~~I={3}/{2}:~
\cO_9&=&(\sbar d)_L [(\ubar u)_R-(\dbar d)_R] + (\sbar u)_L (\ubar d)_R\ \, ,
\nonumber\\
I={3}/{2}:~
\cO_{10}&=&(\sbar_\alpha d_\beta)_L [(\ubar_\beta u_\alpha )_R-
(\dbar_\beta d_\alpha )_R]  + (\sbar_\alpha u_\beta)_L 
(\ubar_\beta d_\alpha )_R \, .\nonumber
\ea
The operators $\cO_5$ and $\cO_6$ are the LR penguin operators considered in 
Ref.~\cite{gp2}. A set of $\Delta S=1$ four-quark effective operators 
frequently used in phenomenological analyses of kaon decays are 
the $Q_i,\,i=1,10$. They are related to our basis as follows:
\ba
\label{Qops}
Q_1&=&\frac{1}{2}\cO_1+\frac{1}{10}\cO_2+\frac{1}{15}\cO_3+\frac{1}{3}\cO_4\, 
,\nonumber\\
Q_2&=&-\frac{1}{2}\cO_1+\frac{1}{10}\cO_2+\frac{1}{15}\cO_3+\frac{1}{3}\cO_4\, ,\nonumber\\
Q_3&=&\frac{1}{2}\cO_1+\frac{1}{2}\cO_2\, ,\nonumber\\
Q_4&=&-\frac{1}{2}\cO_1+\frac{1}{2}\cO_2\, ,\nonumber\\
Q_5&=&\cO_5\, ,~~~Q_6=\cO_6\, ,\nonumber\\
Q_7&=&  \frac{1}{2}(\cO_7+\cO_9)\, ,~~~Q_8=\frac{1}{2}(\cO_8+\cO_{10})\, ,\nonumber\\
Q_9&=&\frac{1}{2}\cO_1-\frac{1}{10}\cO_2+\frac{1}{10}\cO_3+\frac{1}{2}\cO_4 \, ,\nonumber\\
Q_{10}&=&-\frac{1}{2}\cO_1-\frac{1}{10}\cO_2+\frac{1}{10}\cO_3+\frac{1}{2}\cO_4 \, ,
\ea
where we have used $Q_4= Q_2+Q_3-Q_1$, $Q_9=3/2\, Q_1-1/2\, Q_3$ and 
$Q_{10}=3/2\, Q_2-1/2\, Q_4=Q_2-1/2\, Q_3+1/2\, Q_1$.
It appears that any modification induced by partial quenching of the LL 
penguin operators $\cO_{1,2}$ considered in this paper will affect the 
operators $Q_{1,2}$, $Q_{3,4}$ and $Q_{9,10}$ according to the decomposition 
in Eq.~(\ref{Qops}). 
The change of the current-current operators $Q_{1,2}$ is relevant to any 
(partially) quenched lattice calculation of the $\Delta I=1/2$ rule.
A change of the LR penguin operator $Q_6$ can drastically effect a (partially) 
quenched lattice calculation of $\varepsilon '/\varepsilon$ as has been 
analyzed in Ref.~\cite{gp2}. The fact that $Q_4$ is affected as well may also  
be relevant for the determination of $\varepsilon '/\varepsilon$.

\section*{\large\bf Appendix B: Graded groups}
\secteq{B}
In this appendix, we give a few details of graded-group technology as applied
to the PQ operators considered in this paper.

First, our definition of the supertrace of a matrix $M_i^{\ j}$ is
\be
\label{supertrace}
\str(M) = -\sum_i (-1)^{g(i)}M_i^{\ i}\ ,
\ee
where $g(i)=1$ if the index $i$ is fermionic (corresponding to a valence or
sea quark), and $g(i)=0$ if the index $i$ is bosonic (corresponding to a
ghost quark).  The extra overall minus sign is there to make the supertrace
reduce to the normal trace in flavor space in the case of unquenched QCD.

Next, we derive in some detail our claim that the operators $\cO^{PQS}_\pm$
and $\cO^{PQA}_\pm$ are already symmetric or antisymmetric in the pairs of
flavor indices of the quarks (which we refer to as covariant, following 
Ref.~\cite{bb}) and anti-quarks (which we refer to as contravariant).

Begin with considering the operator
\be
\label{op}
(\qbar_{\alpha a}\Lambda q'_{\alpha a})(\qbar_{\beta b}Aq'_{\beta b})
=(\sbar_{\alpha a} d'_{\alpha a})(\qbar_{\beta b}Aq'_{\beta b})\, 
\ee
in which $\alpha$, $\beta$ are color indices and $a$, $b$ are Dirac
indices with summation convention for all explicit indices, and we abbreviate
\be
\label{abbr}
q'=\gamma_\mu P_L q
\ee
for fixed $\mu$.  Symmetrizing this in both covariant and contravariant flavor
indices leads to
\ba
\label{symm}
&&\hspace{-1.5cm}(\qbar_{\alpha a}\Lambda q'_{\alpha a})(\qbar_{\beta b}Aq'_{\beta b})
+(\qbar_{\alpha a}\Lambda q'_{\beta b})(\qbar_{\beta b}Aq'_{\alpha a})
+(\qbar_{\beta b}\Lambda q'_{\alpha a})(\qbar_{\alpha a}Aq'_{\beta b})
+(\qbar_{\beta b}\Lambda q'_{\beta b})(\qbar_{\alpha a}Aq'_{\alpha a})
\nonumber\\
&&\hspace{2cm}=2(\qbar_{\alpha a}\Lambda q'_{\alpha a})(\qbar_{\beta b}Aq'_{\beta b})
+2(\qbar_{\alpha a}\Lambda q'_{\beta b})(\qbar_{\beta b}Aq'_{\alpha a})
\ .
\ea
We used that the symmetric product of two quark fields is given by \cite{bb}
\be
\label{symmquark}
q_{1i} q_{2j}+q_{2i} q_{1j} = q_{1i} q_{2j}+(-1)^{g(i)g(j)}q_{1j} q_{2i}\ ,
\ee
and likewise for anti-quark fields and anti-symmetric products.  
We now wish to write this in a form
in which we can use the shorthand (\ref{shorthand}).  We
rewrite the second term on the right-hand side of Eq.~(\ref{symm}) as
\ba
\label{rewrite}
(\qbar_{\alpha a}\Lambda q'_{\beta b})(\qbar_{\beta b}Aq'_{\alpha a})
&=&\sum_i(-1)^{g(i)}(\sbar_\alpha q_{\alpha i})_L(\qbar_\beta^i d_\beta)_L A_i^{\ i}\\
&=&-(\qbar_\alpha\Lambda q_\beta)_L(\qbar_\beta A q_\alpha)_L\ ,\nonumber
\ea
where in the last step we fierzed the operator, taking into account that
this involves an extra minus sign if $q_{\alpha i}$ and $\qbar_\beta^i$
are bosonic (\ie\ ghosts).  We conclude that symmetrizing the operator of
Eq.~(\ref{op}) in both quark and anti-quark flavor indices yields the operator
$\cO^{PQA}_-$.

A similar argument shows that anti-symmetrizing in both quark and
anti-quark flavor indices leads to the operator $\cO^{PQA}_+$, and that
operators with mixed symmetry (symmetric in quark flavor indices and
anti-symmetric in anti-quark flavor indices, or {\it vice versa}) vanish.
The operators $\cO^{PQS}_\pm$ and $\cO^{PQT}_\pm$ are obtained
in the same way.  Note that these last two pairs of operators are in 
principle different, even though for the particular value of the spurion
field $A$ relevant for this paper (\cf\ Eq.~(\ref{spurions})) we have that
$\cO^{PQT}_\pm=\left(1-\frac{K}{N}\right)\cO^{PQS}_\pm$.

We may now bosonize these four-quark operators following standard
techniques, see for example Ref.~\cite{dghb}.  The only difference with respect to the 
usual, non-graded case is that
care has to be taken with extra signs due to the grading of our symmetry
group, as discussed briefly above, and in much more detail in Ref.~\cite{bb}.
This leads to the weak lagrangians given in Eqs.~(\ref{Lsinglet}) and
(\ref{bosonization}).  In particular, bosonization of $\cO^{PQT}_\pm$ leads
to $\cL^A_3$.  In fact, the effective operators corresponding to $\cO^{PQT}_\pm$
are
\be
\label{L3}
\str(\Lambda AL_\mu L_\mu)\pm\str(\Lambda AL_\mu)\;\str(L_\mu)\ ,
\ee
but we have that $\str(L_\mu)=0$ in the PQ theory with the $\eta'$
integrated out, leading to $\cL^A_3$ in Eq.~(\ref{bosonization}).
In the quenched theory the spurion $A$ is replaced
by $\Nh$, and the $\eta'$ cannot be integrated out, whence $\cL^A_3$
of Eq.~(\ref{bosonization}) is replaced by the quenched version
\be
\label{L3q}
\str(\Lambda \Nh L_\mu L_\mu)\pm\str(\Lambda \Nh L_\mu)\;\str(L_\mu)\ .
\ee


\begin{thebibliography}{99}

\bibitem{review}
  C.~Dawson,
  %``progress in kaon phenomenology from lattice QCD,''
  PoS {\bf LAT2005}, 007 (2005).
  %%CITATION = POSCI,LAT2005,007;%%

\bibitem{cppacs}
  J.~I.~Noaki {\it et al.}  [CP-PACS Collaboration],
  %``Calculation of non-leptonic kaon decay amplitudes from K $\to$ pi matrix
  %elements in quenched domain-wall QCD,''
  Phys.\ Rev.\ D {\bf 68}, 014501 (2003)
  [arXiv:hep-lat/0108013].
  %%CITATION = HEP-LAT 0108013;%%

\bibitem{rbc}
  T.~Blum {\it et al.}  [RBC Collaboration],
  %``Kaon matrix elements and CP-violation from quenched lattice QCD. I: The
  %3-flavor case,''
  Phys.\ Rev.\ D {\bf 68}, 114506 (2003)
  [arXiv:hep-lat/0110075].
  %%CITATION = HEP-LAT 0110075;%%

\bibitem{dghb} For reviews, see
  J.~F.~Donoghue, E.~Golowich and B.~R.~Holstein,
  %``Dynamics of the standard model,''
  Camb.\ Monogr.\ Part.\ Phys.\ Nucl.\ Phys.\ Cosmol.\  {\bf 2}, 1 (1992);
  %%CITATION = CMPCE,2,1;%%
  C.~W.~Bernard,
  %``Weak Matrix Elements On And Off The Lattice,''
  NSF-ITP-89-152
  %\href{http://www.slac.stanford.edu/spires/find/hep/www?r=nsf-itp-89-152}{SPIRES entry}
  {\it Lectures given at TASI '89, Boulder, CO, Jun 4-30, 1989.}
  
\bibitem{bgpq}
C.~W.~Bernard and M.~Golterman,
%``Partially quenched gauge theories and an application to staggered fermions,''
Phys.\ Rev.\ D {\bf 49}, 486 (1994)
[arXiv:hep-lat/9306005].
%%CITATION = HEP-LAT 9306005;%%

\bibitem{gp2}
M.~Golterman and E.~Pallante,
  %``Effects of quenching and partial quenching on penguin matrix elements,''
  JHEP {\bf 0110}, 037 (2001)
  [arXiv:hep-lat/0108010].
  %%CITATION = HEP-LAT 0108010;%%

\bibitem{gp1}
M.~Golterman and E.~Pallante,
%``On the determination of nonleptonic kaon decays from K $\to$ pi matrix
%elements,''
JHEP {\bf 0008}, 023 (2000)
[arXiv:hep-lat/0006029].
%%CITATION = HEP-LAT 0006029;%%

\bibitem{dgs}
P.~H.~Damgaard, J.~C.~Osborn, D.~Toublan and J.~J.~Verbaarschot,
%``The microscopic spectral density of the {QCD} Dirac operator,''
Nucl.\ Phys.\ B {\bf 547}, 305 (1999)
[arXiv:hep-th/9811212];
%%CITATION = HEP-TH 9811212;%%
M.~Golterman, S.~R.~Sharpe and R.~J.~Singleton, Jr., 
  %``Effective theory for quenched lattice QCD and the Aoki phase,''
  Phys.\ Rev.\ D {\bf 71}, 094503 (2005)
  [arXiv:hep-lat/0501015].
  %%CITATION = HEP-LAT 0501015;%%
  
\bibitem{shsh}
S.~R.~Sharpe and N.~Shoresh,
%``Partially quenched chiral perturbation theory without Phi0,''
Phys.\ Rev.\ D {\bf 64}, 114510 (2001)
[arXiv:hep-lat/0108003].
%%CITATION = HEP-LAT 0108003;%%

\bibitem{bb}
A.~Baha Balantekin and I.~Bars,
  %``Dimension And Character Formulas For Lie Supergroups,''
  J.\ Math.\ Phys.\  {\bf 22}, 1149 (1981);
  %%CITATION = JMAPA,22,1149;%%
%``Representations Of Supergroups,''
  1810 (1981);
  %%CITATION = JMAPA,22,1810;%%
{\bf 23}, 1239 (1982).
  %%CITATION = JMAPA,23,1239;%%

\bibitem{gl}
J.~Gasser and H.~Leutwyler,
%``Chiral Perturbation Theory: Expansions In The Mass Of The Strange Quark,''
Nucl.\ Phys.\ B {\bf 250}, 465 (1985).
%%CITATION = NUPHA,B250,465;%%

\bibitem{cbetal}
C.~W.~Bernard, T.~Draper, A.~Soni, H.~D.~Politzer and M.~B.~Wise,
%``Application Of Chiral Perturbation Theory To K $\to$ 2 Pi Decays,''
Phys.\ Rev.\ D {\bf 32} (1985) 2343.    
%%CITATION = PHRVA,D32,2343;%%

\bibitem{bnrs}
M.~Scheunert, W.~Nahm and V.~Rittenberg,
  %``Irreducible Representations Of The Osp(2,1) And Spl(2,1) Graded Lie
  %Algebras,''
  J.\ Math.\ Phys.\  {\bf 18}, 155 (1977);
  %%CITATION = JMAPA,18,155;%%
I.~Bars, B.~Morel and H.~Ruegg,
  %``Kac-Dynkin Diagrams And Supertableaux,''
  J.\ Math.\ Phys.\  {\bf 24}, 2253 (1983).
  %%CITATION = JMAPA,24,2253;%%

\bibitem{shshphys}
  S.~R.~Sharpe and N.~Shoresh,
  %``Physical results from unphysical simulations,''
  Phys.\ Rev.\ D {\bf 62}, 094503 (2000)
  [arXiv:hep-lat/0006017].
  %%CITATION = HEP-LAT 0006017;%%

\bibitem{ls}
  J.~Laiho and A.~Soni,
  %``Lattice extraction of K $\to$ pi pi amplitudes to NLO in partially
  %quenched and in full chiral perturbation theory,''
  Phys.\ Rev.\ D {\bf 71}, 014021 (2005)
  [arXiv:hep-lat/0306035].
  %%CITATION = HEP-LAT 0306035;%%
      
\bibitem{sh}
S.~R.~Sharpe,
  %``Enhanced chiral logarithms in partially quenched QCD,''
  Phys.\ Rev.\ D {\bf 56}, 7052 (1997)
  [Erratum-ibid.\ D {\bf 62}, 099901 (2000)]
  [arXiv:hep-lat/9707018].
  %%CITATION = HEP-LAT 9707018;%%

\bibitem{gl1}
M.~Golterman and K.-C.~Leung,
  %``Applications of partially quenched chiral perturbation theory,''
  Phys.\ Rev.\ D {\bf 57}, 5703 (1998)
  [arXiv:hep-lat/9711033].
  %%CITATION = HEP-LAT 9711033;%%
  
\bibitem{bgq}
  C.~W.~Bernard and M.~Golterman,
  %``Chiral perturbation theory for the quenched approximation of QCD,''
  Phys.\ Rev.\ D {\bf 46}, 853 (1992)
  [arXiv:hep-lat/9204007].
  %%CITATION = HEP-LAT 9204007;%%

\bibitem{sh92}
  S.~R.~Sharpe,
  %``Quenched chiral logarithms,''
  Phys.\ Rev.\ D {\bf 46}, 3146 (1992)
  [arXiv:hep-lat/9205020].
  %%CITATION = HEP-LAT 9205020;%%

\bibitem{Laiho}
C.~Aubin, {\em et al.}, arXiv:hep-lat/0603025. 

\bibitem{Japan}
  T.~Bhattacharya {\em et al.}, Nucl.\ Phys.\ Proc.\ Suppl.\ B {\bf 140}, 
369 (2005)
[arXiv:hep-lat/0409046].

\bibitem{buras}
%'Weak decays beyond leading logarithms.
G.~ Buchalla, A.J.~Buras, M.E.~Lautenbacher, Rev.\ Mod.\ Phys.\ {\bf 68},
1125 (1996) 
[arXiv:hep-ph/9512380]. 




\end{thebibliography}
\end{document}